\documentclass[fleqn,usenatbib]{mnras}
\usepackage{amsmath,amssymb}
\usepackage[T1]{fontenc}
\usepackage{graphicx}
\usepackage{amsmath,amssymb}
\usepackage{listings}
\usepackage{textcomp}
\usepackage{tabularx}
\usepackage{array}
\usepackage{url}

\providecommand{\farcs}{\ensuremath{^{\prime\prime}}}

\DeclareRobustCommand{\VAN}[3]{#2}
\let\VANthebibliography\thebibliography
\def\thebibliography{\DeclareRobustCommand{\VAN}[3]{##3}\VANthebibliography}

\title[PyMsOfa 2.0: Pure-Python IAU SOFA]{PyMsOfa 2.0: A Pure-Python Implementation of the IAU Standards of Fundamental Astronomy Service}

\author[Dongjie Tan et al.]{
Dongjie Tan,$^{1,2}$
Jianghui Ji,$^{1,2,3}$\thanks{E-mail: jijh@pmo.ac.cn (JJ)}
and J.-C. Liu$^{4}$
\\
$^{1}$CAS Key Laboratory of Planetary Sciences, Purple Mountain Observatory, Chinese Academy of Sciences, Nanjing 210023, China\\
$^{2}$School of Astronomy and Space Science, University of Science and Technology of China, Hefei 230026, China\\
$^{3}$CAS Center for Excellence in Comparative Planetology, Hefei 230026, China\\
$^{4}$School of Astronomy and Space Science, Key Laboratory of Modern Astronomy and Astrophysics (Ministry of Education), Nanjing University,\\
163 Xianlin Avenue, Nanjing 210023, China
}

\date{Accepted XXX. Received YYY; in original form ZZZ}
\pubyear{2026}

\begin{document}
\label{firstpage}
\pagerange{\pageref{firstpage}--\pageref{lastpage}}
\maketitle

\begin{abstract}
Astrometric calculations require authoritative algorithms so that every result is correct and reproducible. The Standards of Fundamental Astronomy (SOFA) service of the International Astronomical Union (IAU) provides such algorithms, officially released only in Fortran 77 and ANSI C. In an earlier paper we presented the first version of PyMsOfa, a Python package wrapping the SOFA C library. Here we present PyMsOfa 2.0, which re-implements the complete SOFA library in pure Python with NumPy. All 247 official SOFA routines are ported directly to Python, requiring no compilation of any external C library and running identically on Linux, Windows, and macOS. The routines operate on NumPy arrays, so arguments can be processed in batch, which is convenient for survey pipelines and Monte Carlo simulations. The package also provides an extension module with validated models not yet part of the official SOFA release, such as the improved precession--nutation model IAU 2006$_{J_2}$/2000A$_{R26}$. The reimplementation reproduces the official SOFA algorithms: all constants, series, and coefficients are taken unchanged from the SOFA source files, and routine names differ only by the \texttt{pym} prefix. We verify every routine against the official SOFA C release: most reproduce the C results to machine precision, the remainder within the accuracy tolerance stated in the SOFA documentation. The package has been applied to Gaia astrometric reduction and used in astrometric studies of the Closeby Habitable Exoplanet Survey (CHES) mission, and is available through PyPI, GitHub, and Zenodo.
\end{abstract}

\begin{keywords}
astrometry -- reference systems -- methods: numerical -- software: public release
\end{keywords}

\section{Introduction}
\label{sect:intro}

Astrometry relies on authoritative algorithms: the correctness of a result, and the consistency of the same calculation performed at different times or by different groups, both depend on using algorithms that are agreed upon and precisely defined. The Standards of Fundamental Astronomy (SOFA)\footnote{http://www.iausofa.org/index.html} service of the International Astronomical Union (IAU) provides exactly such a set of algorithms \citep{hohenkerk2011}. SOFA implements the IAU resolutions in authoritative algorithms and software for fundamental astronomy, covering time scales and calendars, coordinate-system transformations, Earth rotation, precession and nutation, and astrometric parameters. The service is maintained by the IAU SOFA Board and is updated as the IAU resolutions are amended, with the latest release dated 2023 October 11 \citep{sofa2023}. The SOFA routines are officially released only in Fortran 77 and ANSI C, and no official Python implementation is provided; PyMsOfa is developed to fill this gap by making the complete SOFA service available natively in Python.

Astrometry has now entered an era of high-precision, all-sky surveys. The Gaia mission provides positions, parallaxes, proper motions, and radial velocities for nearly two billion sources at microarcsecond-level precision \citep{gaia2023}, building on the earlier releases \citep{gaia2018}, and the forthcoming Gaia DR4 will further improve the precision and add time-domain astrometry, including astrometric orbits, radial velocities, and variability. These data underpin the detection of exoplanets, the discovery and characterization of stellar-mass black holes, the realization of the celestial reference frame, and the study of stellar dynamics and the Galactic potential. Reducing such data (propagating catalogue positions to the epoch of observation, applying aberration, gravitational light deflection, and precession--nutation, and transforming between reference systems) requires exactly the authoritative algorithms that SOFA provides. Making these algorithms directly and conveniently available in Python, the language in which most of the Gaia-era reduction pipelines are written, is therefore of broad practical value; the scientific Python ecosystem built on NumPy \citep{harris2020} and community packages such as Astropy \citep{astropy2013} provides general-purpose infrastructure, but these packages do not provide the complete set of authoritative SOFA routines; PyMsOfa fills this gap by supplying the complete SOFA service. Since the scientific interpretation of microarcsecond-level data depends on transformations that are themselves free of systematic error, the routines that perform them must reproduce the official SOFA reference to the full numerical precision.

This is the context in which we develop PyMsOfa. In our earlier paper \citep{ji2023} we presented the first version of PyMsOfa, a Python package that wrapped the SOFA C library and described the main categories of routines and their use in the Closeby Habitable Exoplanet Survey (CHES) mission \citep{ji2022}. CHES is a space-borne astrometric mission designed to detect habitable Earth-like planets around nearby solar-type stars by measuring the sub-microarcsecond reflex motion of the host star; its design, the astrometric noise induced by stellar activity, the observation strategy, the retrieval of planetary masses in binary systems, the synergy with direct-imaging missions, and the derivation of planetary parameters from angular separation variations have been studied in a series of papers \citep{tan2022,ji2022,ji2024cjss,bao2024,dong2024twxb,tan2024,huang2025,bao2025,tan2026}. CHES therefore provides a concrete, high-precision application in which the catalogue-to-observation transformation is exercised end to end: in its early numerical simulations, CHES requires the SOFA algorithms, and hence PyMsOfa, to faithfully simulate the motion of the target stars on the celestial sphere and their observed positions, and its simulation pipelines are written in Python, which is the direct motivation for a native Python SOFA implementation. That first version of PyMsOfa followed the SOFA release of 2023 October 11 and covered all 247 routines.

Here we present PyMsOfa 2.0, in which the complete SOFA library is re-implemented in pure Python with NumPy \citep{harris2020}. The main advantage with respect to the C-based first version is that the routines no longer depend on a compiled library, and that they operate on NumPy arrays so that whole batches of arguments can be processed in a single call. Section~\ref{sect:impl} explains this change and its motivation in detail.

The algorithms themselves are authoritative IAU standards rather than new developments (except for the extension module), so the contribution of PyMsOfa 2.0 lies in making them available natively in Python with NumPy vectorization; because the package is a reimplementation of these algorithms, its correctness is judged by one criterion, namely that the Python results agree with the C results. We therefore devote a section to verification and show that this agreement holds to machine precision (Section~\ref{sect:verif}).

Finally, with the growing number of Python users, new or convenience algorithms can be validated against the SOFA reference and then readily added to PyMsOfa, so that the package serves as an open-source, extensible toolkit for astrometric computation; an example, the IAU 2006$_{J_2}$ precession--nutation model, is described in Section~\ref{sect:app-j2}.

The paper is organized as follows. Section~\ref{sect:impl} describes the pure-Python implementation, its difference from the first version, and the organization of the package. Section~\ref{sect:verif} presents the verification against the official SOFA C library. Section~\ref{sect:app} gives application examples that span the four modules of the package, each illustrated with a short code example. Section~\ref{sect:dist} describes the distribution and the structure of the source repository, and Section~\ref{sect:summary} gives a summary.

\section{The Pure-Python (NumPy) Implementation}
\label{sect:impl}

\subsection{Complete reimplementation of the SOFA algorithms}
\label{sect:impl-port}

In version 2.0 every SOFA routine is translated directly into Python, so that the package no longer depends on the external C library required by the first version. Every constant, series, and coefficient is kept identical to the official SOFA source, and no model, formula, or coefficient is modified, so that the package is a complete reimplementation of the official algorithms rather than a re-derivation. In addition to the 247 routines, the package embeds the standard auxiliary data required by the algorithms, including the IAU 2000A nutation series ($678+687$ terms), the leap-second table, and the reference ellipsoid parameters.

\subsection{Interface conventions}
\label{sect:impl-interface}

All routines keep the original SOFA names with the prefix \texttt{pym} (for example, \texttt{pymCal2jd}, \texttt{pymUtctai}, \texttt{pymPnm06a}, and \texttt{pymApci13}), so that the correspondence with the SOFA C and Fortran libraries is immediate, and output quantities returned through pointer arguments in C are returned as Python return values. The arguments, units, and ranges are documented in the docstring of each routine. Because the names differ from the SOFA C names only by the prefix (\texttt{iau} in C, \texttt{pym} in Python), a user familiar with the SOFA Cookbook can translate an existing C or Fortran program to Python mechanically, and vice versa. In version 2.0 the calling convention has also been modernized: invalid input raises a \texttt{ValueError} instead of returning a trailing integer status, and the \texttt{ASTROM} and \texttt{LDBODY} contexts are ordinary Python classes holding NumPy arrays, in place of the plain Python lists used in the pure-Python back end of version 1 (the \texttt{ctypes} and \texttt{cffi} back ends used their own C structures).

\subsection{From C wrapping to pure Python with NumPy}
\label{sect:impl-numpy}

The most important change with respect to the first version concerns the way the routines are implemented. In the first version the routines followed the C interface closely: to reproduce the official C code, the package wrapped the compiled \texttt{libsofa\_c.so} shared library through the foreign-function interfaces \texttt{ctypes} and \texttt{cffi}. That approach had the advantage of running the byte-identical C algorithms, but it required a compiled library, and each call processed a single set of scalar arguments, so that a survey pipeline looping over millions of sources had to issue the same number of function calls.

In version 2.0 the routines are built on NumPy arrays \citep{harris2020}, so that the same function can operate on whole arrays of quantities at once. This vectorization is convenient for batch processing in survey pipelines and Monte Carlo simulations, and it removes the need to compile or ship any C library, so that the package runs identically on Linux, Windows, and macOS. The following example contrasts the two interfaces:

\begin{lstlisting}[caption={Contrast between the scalar interface of version 1 and the vectorized interface of version 2: a whole array of instants is converted from UTC to TAI in a single call.}, label={lst:vec}]
import PyMsOfa as sf
import numpy as np

# Version 1 (C-wrapped): one instant per call
utc1, utc2 = 2451545.0, 0.5
# UTC -> TAI
tai1, tai2 = sf.pymUtctai(utc1, utc2)

# Version 2 (NumPy): a whole array of instants in one call
# one thousand instants
utc1 = 2451545.0 + np.arange(1000)
utc2 = np.zeros_like(utc1)
# vectorized, single call
tai1, tai2 = sf.pymUtctai(utc1, utc2)
\end{lstlisting}

The two versions are drop-in replacements for one another; the difference is purely in how the arguments are handled.

\subsection{Module organization}
\label{sect:impl-modules}

The 247 SOFA routines are organized into four thematic modules according to their function, together with a module holding the SOFA constants and the auxiliary helper functions, and an extension module for additional validated algorithms. Table~\ref{TabMod} lists the modules with representative routines; the astronomical content of each module was described in the first paper \citep{ji2023}, so here we only note how the modules are laid out in the package. The \texttt{PyMsOfa\_basic} module contains the angle, vector, and matrix utilities on which the other modules build; the \texttt{PyMsOfa\_time} module contains the calendar and time-scale conversions; the \texttt{PyMsOfa\_earth\_attitude} module contains Earth rotation, precession, nutation, and the celestial intermediate reference quantities; and the \texttt{PyMsOfa\_astrometry} module contains the astrometric parameters and the star-space-motion and catalogue-to-observed transformations that combine the other three modules, together with the conversions between the FK4, FK5, and Hipparcos catalog frames \citep{aoki1983}. The \texttt{PyMsOfa\_extension} module provides additional algorithms that are not yet part of the official SOFA release, such as the IAU 2006$_{J_2}$/2000A$_{R26}$ precession--nutation model; these algorithms are supplied and validated by their original authors \citep{liu2025,liu2026} (Section~\ref{sect:app-j2}) and are available at \url{https://zenodo.org/records/22654957}. Every routine and every SOFA constant is also exported from the top level of the package, so that \texttt{import PyMsOfa as sf} makes the full API available without importing the sub-modules individually.

\begin{table*}[!t]
\centering
\caption[]{The modules of PyMsOfa 2.0 with representative routines.}
\label{TabMod}
\begin{tabular}{p{5.6cm}p{1.4cm}p{9.6cm}}
\hline
\hline
\textbf{Module} & \textbf{Routines} & \textbf{Representative routines} \\ \hline
\texttt{PyMsOfa\_basic} & 51 & vector/matrix and angle utilities: \texttt{pymS2c}, \texttt{pymC2s}, \texttt{pymAnp}, \texttt{pymRx}, \texttt{pymA2af} \\ \hline
\texttt{PyMsOfa\_time} & 33 & calendars and time scales: \texttt{pymCal2jd}, \texttt{pymJd2cal}, \texttt{pymUtctai}, \texttt{pymTaitt}, \texttt{pymTttdb} \\ \hline
\texttt{PyMsOfa\_earth\_attitude} & 96 & precession, nutation, Earth rotation: \texttt{pymP06e}, \texttt{pymNut06a}, \texttt{pymPnm06a}, \texttt{pymEra00}, \texttt{pymXys06a} \\ \hline
\texttt{PyMsOfa\_astrometry} & 72 & astrometry, ephemerides, geodesy: \texttt{pymApco13}, \texttt{pymPmsafe}, \texttt{pymAtci13}, \texttt{pymAtco13}, \texttt{pymTpxes} \\ \hline
\texttt{PyMsOfa\_extension} & 13 & IAU 2006$_{J_2}$/2000A$_{R26}$ precession--nutation: \texttt{pymP06J2}, \texttt{pymNut00aR26}, \texttt{pymPnm06J2a}, \texttt{pymC2i06J2a}, \texttt{pymXy06J2} \\ \hline
\hline
\end{tabular}
\end{table*}

\section{Verification against the Official SOFA C Library}
\label{sect:verif}

The correctness of the reimplementation is established by comparing every routine of PyMsOfa with the official SOFA C library. The reference values are taken from the official SOFA test program \texttt{t\_sofa\_c.c}, and a case is regarded as consistent when the maximum absolute difference between the PyMsOfa result and the reference value does not exceed $10^{-12}$, the same judgment precision as the SOFA self-test. Table~\ref{Tab1} lists fifteen representative cases covering the four groups of routines (basic calculations, time systems, Earth rotation and precession--nutation, and astrometry), including exact-value routines, full double-precision transformations, and the two CHES examples used in the first paper \citep{ji2023}.

\begin{table*}[!t]
\centering
\caption[]{Verification of PyMsOfa 2.0 against the official SOFA C library (judgment tolerance $10^{-12}$).}
\label{Tab1}
\begin{tabular}{p{4.0cm}p{8.0cm}cc}
\hline
\hline
\textbf{Group} & \textbf{Routines} & \textbf{Cases} & \textbf{Max $|d|_{\rm max}$} \\ \hline
Basic calculations & \texttt{pymA2af}, \texttt{pymS2c} & 2 & 0 \\ \hline
Time systems & \texttt{pymCal2jd}, \texttt{pymEpj2jd}, \texttt{pymUtctai}, \texttt{pymTaitt} & 4 & 0 \\ \hline
Earth rotation \& precession--nutation & \texttt{pymEra00}, \texttt{pymObl06}, \texttt{pymGmst06}, \texttt{pymPmat06}, \texttt{pymPnm06a} & 5 & $3.8\times10^{-15}$ \\ \hline
Astrometry & \texttt{pymEpv00}, \texttt{pymPmsafe}, \texttt{pymTpxes}, \texttt{pymAtci13} & 4 & 0 \\ \hline
\hline
Total & & 15 & $3.8\times10^{-15}$ \\ \hline
\hline
\end{tabular}
\end{table*}

All fifteen cases are consistent with the SOFA C results: twelve agree bit-for-bit (a maximum difference of exactly zero), and the remaining three differ only in the last one or two digits ($5.6\times10^{-17}$ for the Earth rotation angle and the mean obliquity, and $3.8\times10^{-15}$ for the Greenwich mean sidereal time). These differences are consistent with round-off error and lie far below the $10^{-12}$ tolerance, so the overall agreement is about 15 significant digits, that is, to machine precision.

Beyond these representative cases, a unit-test suite covers all 247 routines and is run automatically on Linux, Windows, and macOS, so that users can reproduce the verification on their own platforms without any compiled dependency. The comparison script is distributed with the package. Together, the routine-by-routine unit tests and the fifteen reference cases confirm that the pure-Python implementation reproduces the authoritative SOFA C algorithms to machine precision.

\section{Applications}
\label{sect:app}

The examples below span the four modules of the package (time, basic, earth-attitude, and astrometry), from the time-scale chain through the catalogue-to-observation transformation to the precession--nutation models. Each example is accompanied by a short, runnable code listing.

\subsection{Time scales and Julian dates}
\label{sect:app-time}

Fundamental to all astrometric work is the handling of time \citep{seidelmann1992}. A civil date is converted to a two-part Julian date with \texttt{pymCal2jd}, in which only the sum of the two parts has meaning,
\begin{equation}
{\rm JD} = {\rm djm0} + {\rm djm},
\end{equation}
a convention that preserves precision when the Julian date is large. The instant can then be carried across the time scales. International Atomic Time is obtained by adding the accumulated leap seconds $\Delta(\mathrm{AT})$,
\begin{equation}
\mathrm{TAI} = \mathrm{UTC} + \Delta(\mathrm{AT}),
\end{equation}
and Terrestrial Time by the fixed offset
\begin{equation}
\mathrm{TT} = \mathrm{TAI} + 32.184~\mathrm{s}.
\end{equation}
Finally, Barycentric Dynamical Time is obtained from TT by the periodic relativistic correction \citep{fairhead1990}
\begin{equation}
\mathrm{TDB}-\mathrm{TT} = 0.001657~\mathrm{s}\cdot\sin g + 0.000022~\mathrm{s}\cdot\sin 2g + \ldots,
\end{equation}
with $g = 357.53^\circ + 0.9856003^\circ\times(\mathrm{JD}-2451545.0)$. The whole chain $\mathrm{UTC}\rightarrow\mathrm{TAI}\rightarrow\mathrm{TT}\rightarrow\mathrm{TDB}$ is realized by \texttt{pymUtctai}, \texttt{pymTaitt}, and \texttt{pymTttdb}, following the conventions of the IERS \citep{mccarthy2004}.

\begin{lstlisting}[caption={A minimal usage example of the PyMsOfa time-scale routines, converting a calendar date to a two-part Julian date and carrying the instant across the UTC, TAI, TT, and TDB time scales.}, label={lst:time}]
import PyMsOfa as sf

# Calendar date to a two-part Julian date (only the SUM of the two parts matters)
djm0, djm = sf.pymCal2jd(2003, 6, 1)

# UTC -> TAI: adds the accumulated leap seconds Delta(AT)
tai1, tai2 = sf.pymUtctai(djm0, djm)

# TAI -> TT: TT = TAI + 32.184 s (exact offset)
tt1, tt2 = sf.pymTaitt(tai1, tai2)

# TT -> TDB: small periodic relativistic correction (|TDB-TT| <= ~1.7 ms)
tdb1, tdb2 = sf.pymTttdb(tt1, tt2, 0.0)

print("JD  :", djm0 + djm)
print("TT  :", tt1 + tt2)
print("TDB :", tdb1 + tdb2)
\end{lstlisting}

\subsection{Stellar positions and the catalogue-to-observation transformation}
\label{sect:app-cat}

Converting a catalogue position into the observed place at the epoch of observation is the first step in many observational tasks, such as computing the apparent place of a star, simulating survey images, and planning telescope pointing. The key to this conversion is the stellar space motion. Taking the Gaia DR3 source $\zeta$~Tuc (catalogue epoch J2016.0) as an example, with ICRS position, proper motion ($\mu_{\alpha*}$, $\mu_\delta$), parallax $\varpi$ and radial velocity $v_r$, the space motion is computed in three dimensions rather than on the sphere. The routine \texttt{pymPmsafe} first combines these catalogue quantities into a Cartesian position--velocity vector, in which the transverse velocity follows from the proper motion and parallax,
\begin{equation}
v_\alpha=\frac{\mu_{\alpha*}}{\varpi}, \qquad v_\delta=\frac{\mu_\delta}{\varpi} \quad (\mathrm{AU~yr}^{-1}),
\end{equation}
and the radial velocity supplies the component along the line of sight. The star's rectilinear space motion is then propagated in three dimensions,
\begin{equation}
r(t)=r_0+v\,\Delta t,
\end{equation}
and the result is converted back to spherical coordinates. Because the transverse velocity is held fixed in linear units while the distance changes, the corresponding angular proper motion is not constant in time; this perspective acceleration is therefore included automatically, without a separate secular correction. The astrometry parameters are then formed by \texttt{pymApco13} (observer position and velocity, light deflection, and aberration \citep{klioner2003,soffel2003}, with the planetary ephemeris of \citet{simon1994}), and the observed place (the apparent right ascension and declination) is obtained with \texttt{pymAtco13} / \texttt{pymAtioq}. With the observed place, one can project it onto the focal plane with \texttt{pymTpxes} through the gnomonic projection
\begin{equation}
\begin{split}
\xi &= \frac{\cos\delta\sin(\alpha-\alpha_0)}{D}, \\
\eta &= \frac{\cos\delta_0\sin\delta-\sin\delta_0\cos\delta\cos(\alpha-\alpha_0)}{D},
\end{split}
\end{equation}
with $D = \sin\delta_0\sin\delta+\cos\delta_0\cos\delta\cos(\alpha-\alpha_0)$, where $(\alpha_0,\delta_0)$ are the direction of the optical axis; or transform it to the horizon (altitude--azimuth) with \texttt{pymGst06a}. This single chain exercises the time, basic, earth-attitude, and astrometry modules together, and is representative of the reduction of ground-based and space-borne observations, including the CHES focal-plane simulations \citep{ji2022,tan2022}.

\begin{lstlisting}[caption={Stellar positions and the catalogue-to-observation transformation. A Gaia DR3 catalogue position is propagated to the epoch of observation by the space motion, transformed to the observed place, and projected onto the tangent plane.}, label={lst:cat}]
import PyMsOfa as sf

# Gaia DR3 catalogue entry (ICRS): position, proper motion, parallax, RV
# position (rad)
ra0, dec0 = 0.08788796644747092, -1.132188403551008
# proper motion (rad/yr)
pmra, pmdec = 8.27454225597756e-06, 5.647882757242327e-06
# parallax (arcsec)
parallax = 116.18257854098105e-3
# radial velocity (km/s)
rv = 9.283571

# (1) Space motion: propagate the catalogue position to the epoch
# catalogue epoch
ep1a, ep1b = 2400000.5, 57388.450625000056
# observation epoch
ep2a, ep2b = 2400000.5, 60991.538933193
ra, dec, pmr, pmd, px, rvr = sf.pymPmsafe(
    ra0, dec0, pmra, pmdec, parallax, rv,
    ep1a, ep1b, ep2a, ep2b)

# (2) ICRS -> CIRS: aberration, light deflection, precession-nutation
ri, di, eo = sf.pymAtci13(
    ra, dec, pmr, pmd, px, rvr,
    2456165.5, 0.401182685)

# (3) Tangent-plane (gnomonic) projection onto the focal plane
xi, eta = sf.pymTpxes(
    0.0902535601897608, -1.1308652476489707,
    ra, dec)

print("Space motion -> ra, dec :", ra, dec)
print("Observed place -> ri, di:", ri, di)
print("Focal plane -> xi, eta  :", xi, eta)
\end{lstlisting}

\begin{figure}[!t]
\centering
\includegraphics[width=\columnwidth]{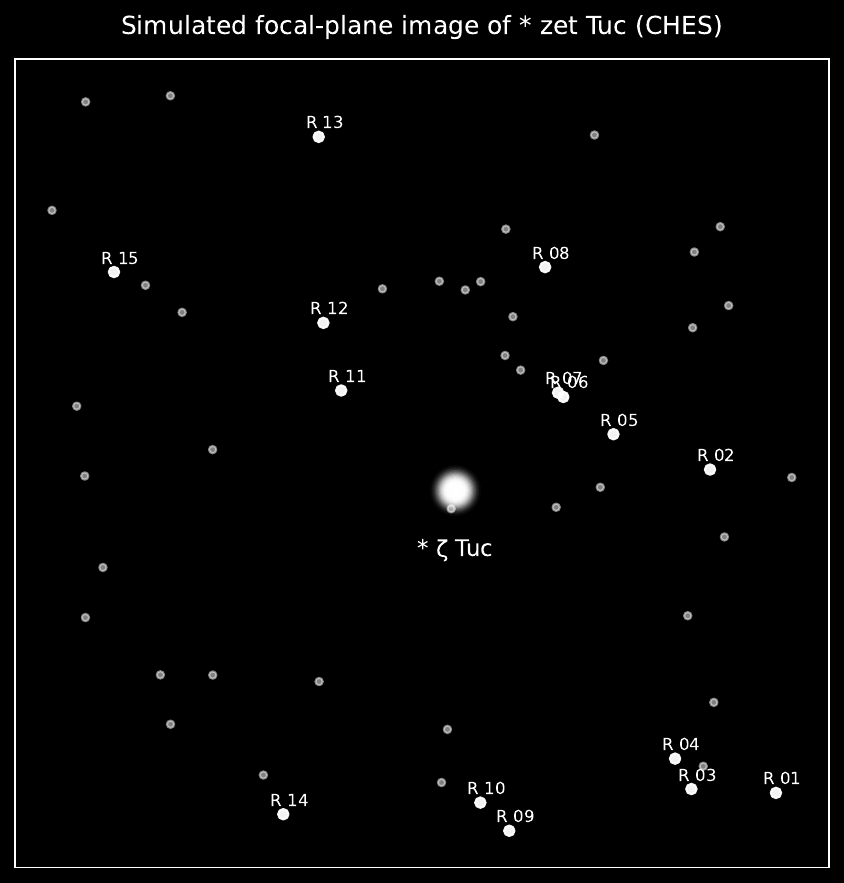}
\caption{Simulated CHES focal-plane image of the target star $\zeta$~Tuc at the observation epoch, obtained by propagating the Gaia DR3 positions to the epoch with \texttt{pymPmsafe} and projecting them onto the tangent plane with \texttt{pymTpxes}.}
\label{Fig1}
\end{figure}

\subsection{Precession-nutation and the celestial-to-observed transformation}
\label{sect:app-pn}

In astrometry, transforming a star's coordinates from the catalogue frame ICRS (which does not rotate with the Earth) into the CIRS (whose reference pole is the celestial intermediate pole and whose origin of right ascension is the celestial intermediate origin) is the common basis of computing the apparent place of a star, planning telescope pointing, reducing observations, and simulating survey images. This transformation is, in essence, the frame bias, precession, and nutation. The IAU 2006 precession angles, accumulated from J2000.0, are polynomial functions of $t$, the TT Julian centuries since J2000.0 \citep{capitaine2003,seidelmann2007}; the general precession in longitude is $p_A \approx 5028.80\farcs\,t + \ldots$. The IAU 2000A nutation is the sum over the luni-solar and planetary terms \citep{mathews2002},
\begin{equation}
\Delta\psi_A = \sum_{j}(S_{j}+S'_{j}t)\sin\phi_{j}, \qquad
\Delta\varepsilon_A = \sum_{j}(C_{j}+C'_{j}t)\cos\phi_{j},
\end{equation}
where each argument $\phi_{j}$ is a linear combination of the fourteen fundamental arguments. This transformation relies on the celestial intermediate pole and celestial intermediate origin, whose definitions and high-precision locations are given by \citet{capitaine2000,capitaine2006}.

The bias--precession--nutation matrix that transforms the Geocentric Celestial Reference System (GCRS) to the Celestial Intermediate Reference System (CIRS) is built from the Fukushima--Williams angles \citep{wallace2006}, with the underlying precession matrix following the IAU precession models of \citet{lieske1979,capitaine2003}:
\begin{equation}
R_{\rm NPB} = R_{1}(-\varepsilon_A-\Delta\varepsilon_A)\cdot R_{3}(-\psi_A-\Delta\psi_A)\cdot R_{1}(\phi)\cdot R_{3}(\gamma).
\end{equation}
For 2025 November 12 (TT, $\mathrm{JD}=2460992.04$) the general precession is $p_A\approx1300.75\farcs\approx0.36^\circ$, and the nutation is $\Delta\psi_A=+3.35\farcs$, $\Delta\varepsilon_A=+8.74\farcs$; the precession rate is about $50.3\farcs$ per year. PyMsOfa performs these calculations with \texttt{pymP06e} (precession angles), \texttt{pymNut06a} (nutation), and \texttt{pymPnm06a} (bias--precession--nutation matrix), as in the code below.

\begin{lstlisting}[caption={Precession and nutation (IAU 2006 / IAU 2000A) computed with PyMsOfa.}, label={lst:pn}]
import PyMsOfa as sf

# Observation epoch (TT), 2025 November 12, as a two-part Julian date
jd = 2460992.04
date1, date2 = jd, 0.0

# Precession angles, IAU 2006 (16 outputs)
(eps0, psia, oma, bpa, bqa, pia, bpia,
 epsa, chia, za, zetaa, thetaa, pa, gam,
 phi, psi) = sf.pymP06e(date1, date2)

# Nutation in longitude and obliquity, IAU 2006/2000A
dpsi, deps = sf.pymNut06a(date1, date2)

# Bias-precession-nutation matrix (GCRS -> CIRS)
rnpb = sf.pymPnm06a(date1, date2)

print("General precession pA (arcsec):", pa / 4.84813681109536e-6)
print("Nutation dpsi, deps (arcsec)  :", dpsi / 4.84813681109536e-6,
      deps / 4.84813681109536e-6)
\end{lstlisting}

\begin{figure*}[!t]
\centering
\includegraphics[width=\textwidth]{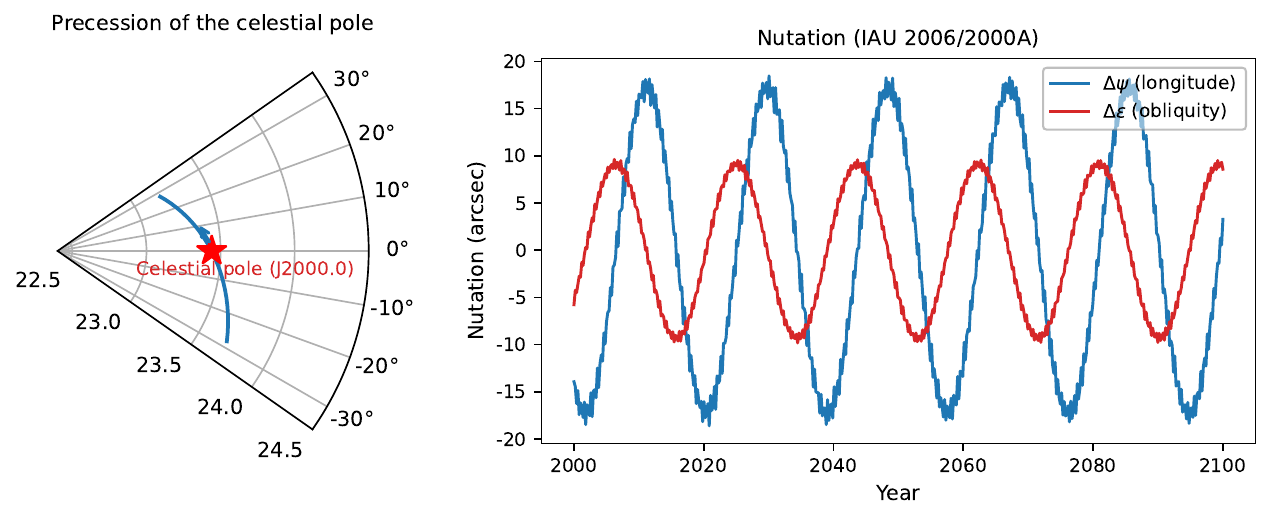}
\caption{Precession and nutation computed with PyMsOfa. Left: the precession arc of the celestial pole about the ecliptic pole, computed with \texttt{pymP06e} over J2000.0 $\pm$ 2000 yr and shown in polar coordinates. Right: the nutation in longitude $\Delta\psi_A$ and obliquity $\Delta\varepsilon_A$ over 2000--2100 from the IAU 2006/2000A model, computed with \texttt{pymNut06a}.}
\label{Fig2}
\end{figure*}

\subsection{Incorporating new models and the IAU 2006$_{J_2}$ precession-nutation}
\label{sect:app-j2}

An important use of PyMsOfa is as an open framework into which validated algorithms that have not yet been formally adopted by the IAU can be incorporated in advance, so that researchers can conveniently use these models, already verified by their proposers, before the IAU resolution is updated. The IAU 2006$_{J_2}$ precession model is given below as an example.

The current IAU 2006 precession \citep{capitaine2003} adopts a linear trend of the Earth's dynamical flattening $J_2$ in solving the precession of the equator (with a rate of about $-3.001\times10^{-9}$ per century), whose relative uncertainty of about 20\% is a main factor limiting the long-term accuracy of the model.

Liu \& Huang \citeyearpar{liu2025} modeled the long-term $J_2$ variation with a parabola using the latest satellite laser ranging (SLR) data, which is 13 years longer than before, and updated the precession quantities of the equator accordingly, naming the new solution IAU 2006$_{J_2}$ (a name given by the proposers, not yet formally adopted by the IAU). Compared with the original IAU 2006 model, the difference lies mainly in the higher-order terms of the precession polynomials, and the advantage is that it is clearly more consistent with the VLBI-observed celestial pole offsets (CPO) and removes most of the curvature signals in the CPO series.

Liu et al.\ \citeyearpar{liu2026} further derived the complete set of operational parameters compatible with this model. They also derived the modified nutation series IAU 2000A$_{R26}$, which adds the $J_2$-dependent corrections and the previously neglected planetary Oppolzer terms \citep{ferrandiz2018} to the 678 luni-solar and 687 planetary terms of the standard IAU 2000A series, together with the celestial intermediate pole coordinates ($X$, $Y$), the CIO locator $s$, the Greenwich sidereal time GST, the Greenwich mean sidereal time GMST, and the equation of the origins EO.

\begin{figure*}[!t]
\centering
\includegraphics[width=\textwidth]{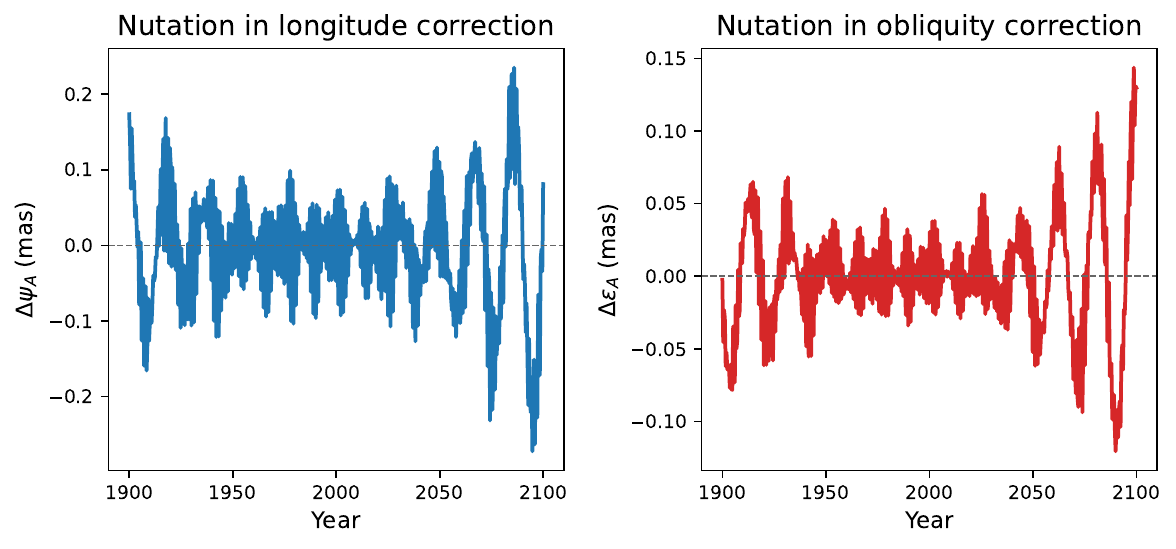}
\caption{Correction to the IAU 2006/2000A nutation introduced by the IAU 2000A$_{R26}$ model, computed with PyMsOfa. Left: the correction to the nutation in longitude $\Delta\psi_A$ as a function of epoch. Right: the correction to the nutation in obliquity $\Delta\varepsilon_A$.}
\label{Fig3}
\end{figure*}

PyMsOfa provides this new model in the extension module \texttt{PyMsOfa\_extension}, whose routines include \texttt{pymP06J2}, \texttt{pymObl06J2}, \texttt{pymPfw06J2}, \texttt{pymNut00aR26}, \texttt{pymXy06J2}, \texttt{pymPnm06J2a}, and \texttt{pymC2i06J2a}, which follow the same \texttt{pym} naming and calling conventions as the rest of the package. Figure~\ref{Fig3} shows the correction that the IAU 2000A$_{R26}$ nutation series introduces to the standard IAU 2006/2000A nutation in longitude and obliquity, computed with PyMsOfa; this correction is produced by the $J_2$-dependent terms and the previously neglected planetary Oppolzer terms.

\begin{lstlisting}[caption={The updated IAU 2006$_{J_2}$ precession-nutation model, provided in the PyMsOfa extension module.}, label={lst:j2}]
import PyMsOfa as sf

# epoch (TT), two-part Julian date
date1, date2 = 2460992.04, 0.0

# IAU 2006J2 precession quantities: psia, oma, pa, epsa, chia
psia, oma, pa, epsa, chia = sf.pymP06J2(date1, date2)

# IAU 2000AR26 nutation in longitude and obliquity
dpsi, deps = sf.pymNut00aR26(date1, date2)

# Bias-precession-nutation and celestial-to-intermediate matrices
rnpb = sf.pymPnm06J2a(date1, date2)
rc2i = sf.pymC2i06J2a(date1, date2)

print("IAU 2006J2 general precession pA (arcsec):", pa / 4.84813681109536e-6)
print("IAU 2000AR26 nutation dpsi, deps (arcsec):",
      dpsi / 4.84813681109536e-6, deps / 4.84813681109536e-6)
\end{lstlisting}

\section{Distribution and Availability}
\label{sect:dist}

The package is distributed through the Python Package Index and can be installed with

\begin{lstlisting}
pip install PyMsOfa
\end{lstlisting}

\noindent The only dependency is NumPy \citep{harris2020}. The source code is hosted on GitHub\footnote{https://github.com/CHES2023/PyMsOfa}. The repository contains the pure-Python package itself, organized into the five modules described in Section~\ref{sect:impl-modules} (\texttt{PyMsOfa\_basic}, \texttt{PyMsOfa\_time}, \texttt{PyMsOfa\_earth\_attitude}, \texttt{PyMsOfa\_astrometry}, and \texttt{PyMsOfa\_extension}), together with the module \texttt{sofa\_const} holding the SOFA constants and helpers, the data module \texttt{iau2006j2\_data.py} implementing the IAU 2006$_{J_2}$/2000A$_{R26}$ precession--nutation model. It also contains the embedded auxiliary data (nutation series, leap-second table, and reference ellipsoid), the unit-test suite and the comparison script used for the verification of Section~\ref{sect:verif}, and the example programs reproduced in Section~\ref{sect:app}.

The algorithms in the extension module are not part of the official SOFA release; they are supplied and validated by their original authors \citep{liu2025,liu2026}. The earlier, C-wrapped version of the package (PyMsOfa 1.x, which followed the SOFA release of 2023 October 11 and used the \texttt{ctypes}/\texttt{cffi} back ends together with the compiled C source) is retained in the repository under the \texttt{v1} directory for reference, so that users who need to reproduce the results of the first paper \citep{ji2023}, or who prefer the compiled C implementation, can still access it. The package layout is shown in Listing~\ref{lst:tree}.

\begin{lstlisting}[caption={Directory structure of the PyMsOfa 2.0 repository.}, label={lst:tree}, language={}]
PyMsOfa/
|-- __init__.py                # flat API
|-- sofa_const.py              # SOFA constants
|-- PyMsOfa_basic.py           # basic (51)
|-- PyMsOfa_time.py            # time (33)
|-- PyMsOfa_earth_attitude.py  # Earth attitude (96)
|-- PyMsOfa_astrometry.py      # astrometry (72)
|-- PyMsOfa_extension.py       # J2 / AR26 (13)
`-- iau2006j2_data.py          # J2 / AR26 data
v1/                            # archived 1.x
examples/                      # examples
docs/                          # Cookbook
.github/workflows/             # CI
\end{lstlisting}

A citable DOI is provided through Zenodo (record 23055828). PyMsOfa is an independent package and is not part of the official SOFA distribution; as with the SOFA library itself, published work that makes use of PyMsOfa should acknowledge the SOFA software and comply with the SOFA terms of use.

For documentation, SOFA itself provides a Cookbook that can serve as a reference. Together with PyMsOfa, we also provide a PyMsOfa version of the Cookbook, which contains a brief summary of the function of each routine and examples of its inputs and outputs. In this Cookbook, all example routines use the \texttt{pym} prefix, while the remaining part of each name is identical to that in SOFA (the C version uses the \texttt{iau} prefix). The Cookbook is therefore useful not only for PyMsOfa users, but also for users of the original SOFA C and Fortran libraries, who can conveniently look up what SOFA and PyMsOfa can do, which routine to use, and what the input and output format is in PyMsOfa. A short excerpt from the Cookbook is shown in Listing~\ref{lst:cookbook}.

\begin{lstlisting}[caption={A short excerpt from the PyMsOfa Cookbook, showing the per-routine summaries together with a code example and its output.}, label={lst:cookbook}, language={}]
pymCal2jd(iy, im, id) :
    Gregorian Calendar to Julian Date.
pymUtctai(utc1, utc2) :
    Time scale transformation: Coordinated Universal Time,UTC,
    to International Atomic Time, TAI.
pymNut06a(date1, date2) :
    IAU 2000A nutation with adjustments to match the
    IAU 2006 precession.
pymPnm06a(date1, date2) :
    Form the matrix of precession-nutation for a given date
    (including frame bias), equinox based, IAU 2006 precession
    and IAU 2000A nutation models.

>>> import PyMsOfa as sf
>>> sf.pymCal2jd(2024, 2, 27)
(2400000.5, 60367.0)
>>> sf.pymUtctai(2400000.5, 60367.0)
(2400000.5, 60367.00042824074)
>>> sf.pymNut06a(2400000.5, 50124.0)
(3.520755263333626e-05, -4.145326347441321e-05)
>>> sf.pymPnm06a(2400000.5, 50124.0)
[[ 9.99999583e-01  8.37238238e-04  3.63968460e-04]
 [-8.37253335e-04  9.99999649e-01  4.13290687e-05]
 [-3.63933730e-04 -4.16337852e-05  9.99999933e-01]]
\end{lstlisting}

\section{Summary}
\label{sect:summary}

Building on the first version of PyMsOfa, we have released PyMsOfa 2.0, which removes the C dependence of the earlier version and re-implements the complete IAU SOFA service in pure Python with NumPy, based on the latest SOFA release of 2023 October 11. Compared with the first version, which wrapped the compiled C library through \texttt{ctypes}/\texttt{cffi} and processed one argument set per call, the new version is built on NumPy arrays and requires no compilation of any external C library, so that the same function can process whole batches of arguments in a single call and the package runs identically on Linux, Windows, and macOS.

The implementation is a complete reimplementation of the authoritative IAU algorithms: all constants, series, and coefficients are taken unchanged from the SOFA source files, and we verified each routine against the official SOFA C release, showing that the two implementations agree to about 15 significant digits. The package covers the full range of fundamental astrometry, from time-scale conversion and precession--nutation to the catalogue-to-observation transformation, which we demonstrated on Gaia data in the context of the CHES mission \citep{ji2022,bao2024,tan2024,bao2025}. As an open framework, it can also incorporate new validated models such as the IAU 2006$_{J_2}$ precession--nutation before they are formally adopted by the IAU, so that researchers can use them in advance.

The package is freely available through PyPI, GitHub, and Zenodo, together with a PyMsOfa version of the Cookbook, and the earlier C-wrapped version remains in the repository for reference. Future work will keep the package synchronized with new SOFA releases, extend its applications to the forthcoming Gaia data releases and to further astrometric studies, and continue to incorporate newly validated models as they appear.

\section*{Acknowledgements}

This work is financially supported by the National Natural Science Foundation
of China (grant Nos. 12533011, 12033010, 12473076), the Strategic Priority
Research Program on Space Science of the Chinese Academy of Sciences (Grant
No. XDA 15020800), and the Foundation of Minor Planets of the Purple Mountain
Observatory. We thank the IAU SOFA Board for the SOFA software and its
documentation.

\section*{Data Availability}

PyMsOfa 2.0 is released under an open-source licence and is available from
GitHub (\url{https://github.com/CHES2023/PyMsOfa}), the Python Package Index
(\url{https://pypi.org/project/PyMsOfa/}), and Zenodo (record 23055828). The
source repository contains the pure-Python implementation, the embedded
auxiliary data, the unit-test suite and comparison scripts used for the
verification in Section~\ref{sect:verif}, and the example programs reproduced
in Section~\ref{sect:app}. The IAU 2006$_{J_2}$/2000A$_{R26}$ model data are
available at \url{https://zenodo.org/records/22654957}.

\bibliographystyle{mnras}
\bibliography{reference}

@ARTICLE{ji2023,
       author = {{Ji}, Jianghui and {Tan}, Dongjie and {Bao}, Chunhui and {Huang}, Xiumin and {Hu}, Shoucun and {Dong}, Yao and {Wang}, Su},
        title = "{PyMsOfa: A Python Package for the Standards of Fundamental Astronomy (SOFA) Service}",
      journal = {Research in Astronomy and Astrophysics},
         year = 2023,
        month = dec,
       volume = {23},
       number = {12},
          eid = {125015},
        pages = {125015},
          doi = {10.1088/1674-4527/ad0499},
}

@ARTICLE{ji2022,
       author = {{Ji}, Jiang-Hui and {Li}, Hai-Tao and {Zhang}, Jun-Bo and {Fang}, Liang and {Li}, Dong and {Wang}, Su and {Cao}, Yang and {Deng}, Lei and {Li}, Bao-Quan and {Xian}, Hao and {Gao}, Xiao-Dong and {Zhang}, Ang and {Li}, Fei and {Liu}, Jia-Cheng and {Qi}, Zhao-Xiang and {Jin}, Sheng and {Liu}, Ya-Ning and {Chen}, Guo and {Li}, Ming-Tao and {Dong}, Yao and {Zhu}, Zi and {CHES Consortium}},
        title = "{CHES: A Space-borne Astrometric Mission for the Detection of Habitable Planets of the Nearby Solar-type Stars}",
      journal = {Research in Astronomy and Astrophysics},
         year = 2022,
        month = jul,
       volume = {22},
       number = {7},
          eid = {072003},
        pages = {072003},
          doi = {10.1088/1674-4527/ac77e4},
}

@ARTICLE{tan2022,
       author = {{Tan}, Dong-Jie and {Liu}, Jia-Cheng and {Zhu}, Zi and {Liu}, Ni},
        title = "{Evaluating the Impact of Optical Axis Stability on Exoplanet Detection}",
      journal = {Research in Astronomy and Astrophysics},
         year = 2022,
        month = feb,
       volume = {22},
       number = {2},
          eid = {025008},
        pages = {025008},
          doi = {10.1088/1674-4527/ac3df1},
}

@ARTICLE{bao2024,
       author = {{Bao}, Chunhui and {Ji}, Jianghui and {Tan}, Dongjie and {Chen}, Guo and {Huang}, Xiumin and {Wang}, Su and {Dong}, Yao},
        title = "{Closeby Habitable Exoplanet Survey (CHES). I. Astrometric Noise and Planetary Detection Efficiency due to Stellar Spots and Faculae}",
      journal = {The Astronomical Journal},
         year = 2024,
        month = apr,
       volume = {167},
          eid = {286},
        pages = {286},
          doi = {10.3847/1538-3881/ad4031},
}

@ARTICLE{tan2024,
       author = {{Tan}, Dongjie and {Ji}, Jianghui and {Bao}, Chunhui and {Huang}, Xiumin and {Chen}, Guo and {Wang}, Su and {Dong}, Yao and {Li}, Haitao and {Zhang}, Junbo and {Fang}, Liang and {Li}, Dong and {Deng}, Lei and {Liu}, Jiacheng and {Zhu}, Zi},
        title = "{Closeby Habitable Exoplanet Survey (CHES). II. An Observation Strategy for the Target Stars}",
      journal = {The Astronomical Journal},
         year = 2024,
        month = sep,
       volume = {168},
          eid = {172},
        pages = {172},
          doi = {10.3847/1538-3881/ad6f01},
}

@ARTICLE{huang2025,
       author = {{Huang}, Xiumin and {Ji}, Jianghui and {Bao}, Chunhui and {Tan}, Dongjie and {Wang}, Su and {Dong}, Yao and {Chen}, Guo},
        title = "{Closeby Habitable Exoplanet Survey (CHES). III. Retrieval of Planetary Masses in Binaries Using the N-body Model with Radial Velocity and Astrometry Synergy}",
      journal = {The Astrophysical Journal},
         year = 2025,
       volume = {984},
          eid = {82},
        pages = {82},
          doi = {10.3847/1538-4357/adc444},
}

@ARTICLE{bao2025,
       author = {{Bao}, Chunhui and {Ji}, Jianghui and {Tan}, Dongjie and {Chen}, Guo and {Huang}, Xiumin and {Wang}, Su and {Dong}, Yao},
        title = "{Closeby Habitable Exoplanet Survey (CHES). IV. Synergy between Astrometry and Direct Imaging Missions of the Habitable World Observatory for Detecting Earth-like Planets}",
      journal = {The Astronomical Journal},
         year = 2025,
        month = jul,
       volume = {170},
          eid = {26},
        pages = {26},
          doi = {10.3847/1538-3881/add5de},
}

@ARTICLE{tan2026,
       author = {{Tan}, Dongjie and {Ji}, Jianghui and {Bao}, Chunhui and {Huang}, Xiumin and {Chen}, Guo and {Wang}, Su and {Dong}, Yao and {Liu}, Jiacheng and {Zhu}, Zi and {Li}, Haitao and {Zhang}, Junbo and {Fang}, Liang and {Li}, Dong and {Deng}, Lei},
        title = "{Closeby Habitable Exoplanet Survey (CHES). V. Planetary Parameters Derived from Angular Separation Variations}",
      journal = {Research in Astronomy and Astrophysics},
         year = 2026,
       volume = {26},
       number = {8},
          doi = {10.1088/1674-4527/ae5b55},
}

@ARTICLE{ji2024cjss,
       author = {{Ji}, Jianghui and {Li}, Haitao and {Zhang}, Junbo and {Li}, Dong and {Fang}, Liang and {Wang}, Su and {Deng}, Lei and {Chen}, Guo and {Li}, Fei and {Dong}, Yao and {Li}, Baoquan and {Gao}, Xiaodong and {Xian}, Hao},
        title = "{Closeby Habitable Exoplanet Survey (CHES): an Astrometry Mission for Probing Nearby Habitable Planets}",
      journal = {Chinese Journal of Space Science},
         year = 2024,
        month = apr,
       volume = {44},
       number = {2},
        pages = {193},
          doi = {10.11728/cjss2024.02.yg03},
}

@ARTICLE{dong2024twxb,
       author = {{Bao}, Chunhui and {Ji}, Jianghui and {Tan}, Dongjie and {Jin}, Sheng and {Huang}, Xiumin and {Wang}, Su and {Chen}, Guo and {Dong}, Yao and {Liu}, Jiacheng},
        title = "{Observation Simulation and Orbital Retrieval of the CHES Satellite}",
      journal = {Acta Astronomica Sinica},
         year = 2024,
       volume = {65},
       number = {4},
          doi = {10.15940/j.cnki.0001-5245.2024.04.006},
}

@ARTICLE{gaia2023,
       author = {{Gaia Collaboration} and {Vallenari}, A. and {Brown}, A.~G.~A. and {Prusti}, T. and others},
        title = "{Gaia Data Release 3. Summary of the content and survey properties}",
      journal = {Astronomy and Astrophysics},
         year = 2023,
        month = jun,
       volume = {674},
          eid = {A1},
        pages = {A1},
          doi = {10.1051/0004-6361/202243940},
}

@ARTICLE{gaia2018,
       author = {{Gaia Collaboration} and {Brown}, A.~G.~A. and {Vallenari}, A. and {Prusti}, T. and others},
        title = "{Gaia Data Release 2. Summary of the contents and survey properties}",
      journal = {Astronomy and Astrophysics},
         year = 2018,
        month = aug,
       volume = {616},
          eid = {A1},
        pages = {A1},
          doi = {10.1051/0004-6361/201833051},
}

@ARTICLE{harris2020,
       author = {{Harris}, Charles R. and {Millman}, K. Jarrod and {van der Walt}, St{\'e}fan J. and {Gommers}, Ralf and {Virtanen}, Pauli and {Cournapeau}, David and {Wieser}, Eric and {Taylor}, Julian and {Berg}, Sebastian and {Smith}, Nathaniel J. and {Kern}, Robert and {Picus}, Matti and {Hoyer}, Stephan and {van Kerkwijk}, Marten H. and {Brett}, Matthew and {Haldane}, Allan and {del R{\'\i}o}, Jaime Fern{\'a}ndez and {Wiebe}, Mark and {Peterson}, Pearu and {G{\'e}rard-Marchant}, Pierre and {Sheppard}, Kevin and {Reddy}, Tyler and {Weckesser}, Warren and {Abbasi}, Hameer and {Gohlke}, Christoph and {Oliphant}, Travis E.},
        title = "{Array programming with NumPy}",
      journal = {Nature},
         year = 2020,
        month = sep,
       volume = {585},
        pages = {357-362},
          doi = {10.1038/s41586-020-2649-2},
}

@ARTICLE{astropy2013,
       author = {{Astropy Collaboration} and {Robitaille}, Thomas P. and {Tollerud}, Erik J. and others},
        title = "{Astropy: A community Python package for astronomy}",
      journal = {Astronomy and Astrophysics},
         year = 2013,
        month = oct,
       volume = {558},
          eid = {A33},
        pages = {A33},
          doi = {10.1051/0004-6361/201322068},
}

@ARTICLE{hohenkerk2011,
       author = {{Hohenkerk}, C.~Y.},
        title = "{Standards of Fundamental Astronomy}",
      journal = {Scholarpedia},
         year = 2011,
        month = jan,
       volume = {6},
       number = {1},
        pages = {11404},
          doi = {10.4249/scholarpedia.11404},
}

@MISC{sofa2023,
      author = {{IAU SOFA Board}},
       title = "{SOFA: Standards of Fundamental Astronomy, software and documentation}",
        year = 2023,
        howpublished = {\url{http://www.iausofa.org/}},
}

@ARTICLE{capitaine2000,
       author = {{Capitaine}, N. and {Guinot}, B. and {McCarthy}, D.~D.},
        title = "{Definition of the Celestial Ephemeris Origin and of UT1 in the International Celestial Reference Frame}",
      journal = {Astronomy and Astrophysics},
         year = 2000,
        month = mar,
       volume = {355},
        pages = {398-405},
}

@ARTICLE{capitaine2003,
       author = {{Capitaine}, N. and {Wallace}, P.~T. and {Chapront}, J.},
        title = "{Expressions for IAU 2000 precession quantities}",
      journal = {Astronomy and Astrophysics},
         year = 2003,
        month = dec,
       volume = {412},
        pages = {567-586},
          doi = {10.1051/0004-6361:20031539},
}

@ARTICLE{capitaine2006,
       author = {{Capitaine}, N. and {Wallace}, P.~T.},
        title = "{High precision methods for locating the celestial intermediate pole and origin}",
      journal = {Astronomy and Astrophysics},
         year = 2006,
        month = apr,
       volume = {450},
        pages = {855-872},
          doi = {10.1051/0004-6361:20054550},
}

@ARTICLE{wallace2006,
       author = {{Wallace}, P.~T. and {Capitaine}, N.},
        title = "{Precession-nutation procedures consistent with IAU 2006 resolutions}",
      journal = {Astronomy and Astrophysics},
         year = 2006,
        month = nov,
       volume = {459},
        pages = {981-985},
          doi = {10.1051/0004-6361:20065897},
}

@ARTICLE{mathews2002,
       author = {{Mathews}, P.~M. and {Herring}, T.~A. and {Buffett}, B.~A.},
        title = "{Modeling of nutation and precession: New nutation series for nonrigid Earth and insights into the Earth's interior}",
      journal = {Journal of Geophysical Research (Solid Earth)},
         year = 2002,
        month = apr,
       volume = {107},
       number = {B4},
        pages = {2068},
          doi = {10.1029/2001JB000390},
}

@ARTICLE{lieske1979,
       author = {{Lieske}, J.~H.},
        title = "{Precession matrix based on IAU /1976/ system of astronomical constants}",
      journal = {Astronomy and Astrophysics},
         year = 1979,
        month = mar,
       volume = {73},
        pages = {282-284},
}

@ARTICLE{seidelmann2007,
       author = {{Seidelmann}, P.~K. and {Archinal}, B.~A. and {A'Hearn}, M.~F. and {Conrad}, A. and {Consolmagno}, G.~J. and {Hestroffer}, D. and {Hilton}, J.~L. and {Krasinsky}, G.~A. and {Neumann}, G. and {Oberst}, J. and {Stooke}, P. and {Tedesco}, E.~F. and {Tholen}, D.~J. and {Thomas}, P.~C. and {Williams}, I.~P.},
        title = "{Report of the IAU/IAG Working Group on cartographic coordinates and rotational elements: 2006}",
      journal = {Celestial Mechanics and Dynamical Astronomy},
         year = 2007,
        month = jul,
       volume = {98},
        pages = {155-180},
          doi = {10.1007/s10569-007-9072-y},
}

@ARTICLE{ferrandiz2018,
       author = {{Ferr{\'a}ndiz}, J.~M. and {Navarro}, J.~F. and {Mart{\'\i}nez-Belda}, M.~C. and {Escapa}, A. and {Getino}, J.},
        title = "{Limitations of the IAU2000 nutation model accuracy due to the lack of Oppolzer terms of planetary origin}",
      journal = {Astronomy and Astrophysics},
         year = 2018,
        month = oct,
       volume = {618},
          eid = {A69},
        pages = {A69},
          doi = {10.1051/0004-6361/201730840},
}

@ARTICLE{liu2025,
       author = {{Liu}, J.-C. and {Huang}, C.-L.},
        title = "{The IAU 2006 precession quantities with an improved Earth's J2 long-term variation}",
      journal = {Astronomy and Astrophysics},
         year = 2025,
        month = nov,
       volume = {703},
          eid = {L21},
        pages = {L21},
          doi = {10.1051/0004-6361/202557665},
}

@ARTICLE{liu2026,
       author = {{Liu}, J.-C. and {Yao}, J. and {Escapa}, A. and {Huang}, C.-L. and {Ferr{\'a}ndiz}, J.~M. and {Karbon}, M.},
        title = "{Precession-nutation quantities compatible with the IAU 2006J2 precession model}",
      journal = {Astronomy and Astrophysics},
         year = 2026,
          doi = {10.1051/0004-6361/202661238},
}

@BOOK{mccarthy2004,
       author = {{McCarthy}, D.~D. and {Petit}, G.},
        title = "{IERS Conventions (2003)}",
     publisher = {Verlag des Bundesamts f{\"u}r Kartographie und Geod{\"a}sie},
         year = 2004,
      address = {Frankfurt am Main},
}

@ARTICLE{klioner2003,
       author = {{Klioner}, S.~A.},
        title = "{A Practical Relativistic Model for Microarcsecond Astrometry in Space}",
      journal = {The Astronomical Journal},
         year = 2003,
        month = mar,
       volume = {125},
        pages = {1580-1597},
          doi = {10.1086/367593},
}

@ARTICLE{simon1994,
       author = {{Simon}, J.~L. and {Bretagnon}, P. and {Chapront}, J. and {Chapront-Touz{\'e}}, M. and {Francou}, G. and {Laskar}, J.},
        title = "{Numerical expressions for precession formulae and mean elements for the Moon and the planets}",
      journal = {Astronomy and Astrophysics},
         year = 1994,
        month = feb,
       volume = {282},
        pages = {663-683},
}

@ARTICLE{fairhead1990,
       author = {{Fairhead}, L. and {Bretagnon}, P.},
        title = "{An analytical formula for the time transformation TB-TT}",
      journal = {Astronomy and Astrophysics},
         year = 1990,
        month = mar,
       volume = {229},
        pages = {240-247},
}

@ARTICLE{aoki1983,
       author = {{Aoki}, S. and {Soma}, M. and {Kinoshita}, H. and {Inoue}, K.},
        title = "{Conversion matrix of epoch B 1950.0 FK 4-based positions of stars to epoch J 2000.0 positions in accordance with the new IAU resolutions}",
      journal = {Astronomy and Astrophysics},
         year = 1983,
        month = nov,
       volume = {128},
        pages = {263-267},
}

@ARTICLE{soffel2003,
       author = {{Soffel}, M. and {Klioner}, S.~A. and {Petit}, G. and {Wolf}, P. and {Kopeikin}, S.~M. and {Bretagnon}, P. and {Brumberg}, V.~A. and {Capitaine}, N. and {Damour}, T. and {Fukushima}, T. and {Guinot}, B. and {Huang}, T.-Y. and {Lindegren}, L. and {Ma}, C. and {Nordtvedt}, K. and {Ries}, J.~C. and {Seidelmann}, P.~K. and {Vokrouhlick{\'y}}, D. and {Will}, C.~M. and {Xu}, C.},
        title = "{The IAU 2000 Resolutions for Astrometry, Celestial Mechanics, and Metrology in the Relativistic Framework: Explanatory Supplement}",
      journal = {The Astronomical Journal},
         year = 2003,
        month = dec,
       volume = {126},
        pages = {2687-2706},
          doi = {10.1086/378162},
}

@BOOK{seidelmann1992,
       author = {{Seidelmann}, P.~K.},
        title = "{Explanatory Supplement to the Astronomical Almanac}",
     publisher = {University Science Books},
         year = 1992,
      address = {Mill Valley, CA},
}

\bsp
\label{lastpage}
\end{document}